\documentclass[a4paper,conference]{IEEEtran}

\usepackage[T1]{fontenc} 
\usepackage[utf8]{inputenc} 
\usepackage{amsmath,url}
\usepackage{graphicx}
\usepackage{color}
\usepackage[table,x11names]{xcolor}
\definecolor{Gray}{gray}{0.9}

\newcommand\addgithub{https://github.com/KinWaiCheuk/ICPR2020}

\begin{document}
%
\title{\huge The Effect of Spectrogram Reconstruction on \\ Automatic Music Transcription:\\An Alternative Approach to Improve Transcription Accuracy}

\author{
\IEEEauthorblockN{Kin Wai Cheuk\IEEEauthorrefmark{1}$^{1,2}$, Yin-Jyun Luo\IEEEauthorrefmark{2}$^{3}$, Emmanouil Benetos\IEEEauthorrefmark{3}$^{3}$, Dorien Herremans\IEEEauthorrefmark{4}$^{1,2}$}
\IEEEauthorblockA{$^{1}$ Information Systems Technology and Design,
                    Singapore University of Technology and Design\\
                    $^{2}$ Institute of High Performance Computing,
                    Agency for Science, Technology and Research\\
$^{3}$ School of Electronic Engineering and Computer Science, Queen Mary University of London}

\\Email: \IEEEauthorrefmark{1}kinwai\_cheuk@mymail.sutd.edu.sg,\IEEEauthorrefmark{2}yin-jyun.luo@qmul.ac.uk,\\\IEEEauthorrefmark{3}emmanouil.benetos@qmul.ac.uk, \IEEEauthorrefmark{4}dorien\_herremans@sutd.edu.sg}


%


\maketitle

\begin{abstract}
Most of the state-of-the-art automatic music transcription (AMT) models break down the main transcription task into sub-tasks such as onset prediction and offset prediction and train them with onset and offset labels. These predictions are then concatenated together and used as the input to train another model with the pitch labels to obtain the final transcription. We attempt to use only the pitch labels (together with spectrogram reconstruction loss) and explore how far this model can go without introducing supervised sub-tasks. In this paper, we do not aim at achieving state-of-the-art transcription accuracy, instead, we explore the effect that spectrogram reconstruction has on our AMT model. Our proposed model consists of two U-nets: the first U-net transcribes the spectrogram into a posteriorgram, and a second U-net transforms the posteriorgram back into a spectrogram. A reconstruction loss is applied between the original spectrogram and the reconstructed spectrogram to constrain the second U-net to focus only on reconstruction. We train our model on three different datasets: MAPS, MAESTRO, and MusicNet. Our experiments show that adding the reconstruction loss can generally improve the note-level transcription accuracy when compared to the same model without the reconstruction part. Moreover, it can also boost the frame-level precision to be higher than the state-of-the-art models. The feature maps learned by our U-net contain gridlike structures (not present in the baseline model) which implies that with the presence of the reconstruction loss, the model is probably trying to count along both the time and frequency axis, resulting in a higher note-level transcription accuracy.

\end{abstract}

\section{Introduction}\label{sec:introduction}

Automatic music transcription (AMT) is usually defined as the process of transforming audio signals into piano-roll representations~\cite{cemgil2004bayesian}, or score formats~\cite{carvalho2017towards, roman2018end, roman2019holistic}. We focus on the former in this paper. Transcribing audio has a multitude of applications, such as turning a huge set of audio data into an indexable format to enable queries based on musical structure~\cite{cuthbert2010music21}, converting an audio dataset to a symbolic dataset as the input for a symbolic music generation model~\cite{huang2020pop}, or to enable musicians to play along~\cite{magalhaeschordify}.
Various models have been proposed to tackle this task, from older generations of machine learning algorithms such as support vector machines (SVMs)~\cite{Poliner2006ADM,Poliner2007IMPROVINGGF} and restricted Boltzmann machines (RBMs)~\cite{BoulangerLewandowski2012ModelingTD} to deep learning models such as fully convolutional neural networks~\cite{springenberg2014striving}, hybrid convolutional and recurrent neural networks~\cite{Sigtia2015AnEN}, and convolutional sequence to sequence models~\cite{ullrich2018music}. These models use the piano roll as the ground truth and train their model by minimizing a loss, usually the binary cross-entropy, between the predicted posteriorgram and the ground truth. Hawthorne et al.~\cite{Hawthorne2017OnsetsAF} tried to extract onset information from the piano roll, and use individual models to predict onsets and multiple pitches at a frame-level. Then a bidirectional long short-term memory (biLSTM) block takes in both the predicted onsets and pitches and produces the final transcription, in the form of a piano roll. Since then, people have followed this approach to extract more information from the ground truth, and use these handcrafted extra labels to train specialized networks to predict each of them. For example, Kim and Bello~\cite{kim2019adversarial} extract offset information on top of the onsets and pitch labels, and obtain a better transcription result than Hawthorne et al.~\cite{Hawthorne2017OnsetsAF}. Similarly Kelz et al.~\cite{kelz2019deep} also use onsets, offsets, and pitch labels to achieve state-of-the-art performance.

The above approaches are heavily supervised due to the fact that they introduce sub-tasks such as onset and offset predictions. In this paper, we try to explore an alternative approach which exploits the spectrogram reconstruction instead of introducing the above-mentioned sub-tasks. Some unsupervised music transcription models can transcribe music solely from the spectrogram without any help from the ground truth piano roll. For example, Kameoka et al.~\cite{kameoka2007multipitch} use a method called harmonic temporal structured clustering to detect both the musical instruments and the notes being played; Berg-Kirkpatrick et al.~\cite{berg2014unsupervised} use a graphical model which consists of event, envelope and spectral information as the parameters.  Once the spectrogram is being reconstructed successfully using these parameters, the notes being played by the piano can be inferred from these parameters. However, unsupervised neural network-based polyphonic music transcription is still an unexplored area, and the only attempt, to the best of our knowledge, is the unsupervised drum transcription~\cite{DBLP:conf/ismir/ChoiC19}. Therefore, we move towards this research direction by proposing a model that uses spectrogram reconstruction and study its effect on  transcription accuracy. We pay more attention to note-level metrics because they reflect the model performance better at a perceptual level as pointed out by various studies~\cite{Hawthorne2017OnsetsAF, ycart2019blending}. Our proposed model\footnote{\addgithub} consists of two U-nets~\cite{ronneberger2015u} (see~\figurename\ref{fig:model}). The first U-net (U-net 1) acts as the transcriber, which transforms the spectrograms into posteriorgrams, and the second U-net (U-net 2) is the reconstructer, which reconstructs the spectrograms from the posteriorgrams. We apply one BiLSTM (BiLSTM 1 in~\figurename\ref{fig:model}) after U-net 1 and another BiLSTM (BiLSTM 2 in~\figurename\ref{fig:model}) before U-net 2, so as to model inter-frame dependencies. We opted to use U-nets since we treat AMT as a sequence to sequence problem, such that the length of the transcriptions depends on the length of the input spectrograms. Therefore, U-net is the suitable model for this formulation as we will discuss in Section~\ref{sec: formulation}. In Section~\ref{sec:Results}, we will discuss our model performance trained on the MAPS~\cite{emiya2010maps}, MAESTRO~\cite{hawthorne2018enabling}, and MusicNet~\cite{Thickstun2017InvariancesAD} datasets respectively. Although our proposed model is not fully unsupervised, we aim to contribute to the investigation of AMT using the reconstruction loss which is an important component for an unsupervised model. The main contributions for this paper are:
\begin{enumerate}
\setlength\itemsep{0.05em}
  \item Improving note-level transcription accuracy and frame-level precision using only pitch labels.
  \item Including spectrogram reconstruction as an extra task for our transcription model.
  \item Using the L2 reconstruction loss to guide the model to discover onset and offset features without providing explicit labels.
\end{enumerate}

\section{Background}
\subsection{Types of Automatic Transcription models}

We can classify automatic music transcription (AMT) models into two types based on their model input and output. If the model takes in a small chunk of the spectrogram as the input, and predicts the pitches at only that time step, then this model can be considered as a \textbf{frame-based model}; if the model takes a segment as input, or even the full spectrogram, and predicts the sequences of notes within that segment, then it can be considered as a \textbf{note-based model}.  Frame-based models usually have less complexity because they just need to transcribe each spectrogram chunk independently, without considering either the temporal dependencies among the spectrogram chunks or the long term structure of the music. Many of the older models are frame-based~\cite{Poliner2006ADM, Poliner2007IMPROVINGGF, nam2011classification} including some neural network models~\cite{trabelsi2017deep, Thickstun2016LearningFO, Thickstun2017InvariancesAD, pedersoli2020improving}. Note-based models are usually more complicated than frame based models due to the fact that the input spectrograms are longer, and we need to transcribe multiple time steps at once. In order for a note-based model to perform well, it needs to understand the long-term structure of the music. Hence, these types of models usually contain either a hidden Markov model (HMM)~\cite{kelz2019deep} or a recurrent neural network (RNN) to model the long term structures~\cite{Sigtia2015AnEN, bock2012polyphonic, Hawthorne2017OnsetsAF, kim2019adversarial}. Nonetheless, both types of models generate piano rolls as the output.
Our proposed model belongs to the latter type since it is able to transcribe multiple time steps at once depending on the input spectrogram size.

\subsection{Pitches, Onset, and Offset Labels}
Most of the music transcription datasets such as MAPS~\cite{emiya2010maps} and MAESTRO~\cite{hawthorne2018enabling} provide audio recordings, together with corresponding aligned music notations, from which the ground truth piano rolls can be extracted. A piano roll is a binary representation indicating if a pitch is `on' or `off' for each frame. This information can be used as pitch labels. Even when the dataset does not contain MIDI files, they would provide comma-separated values (CSV) files in which the onsets and offsets of the pitches are annotated, e.g., MusicNet~\cite{Thickstun2016LearningFO}. The piano rolls (pitch labels) can also be extracted from these CSV files. Piano rolls have been widely used by various models as the ground truth label, regardless of the model architecture~\cite{Poliner2006ADM, Poliner2007IMPROVINGGF, nam2011classification,trabelsi2017deep, Thickstun2016LearningFO, Thickstun2017InvariancesAD, pedersoli2020improving, Sigtia2015AnEN, bock2012polyphonic, Hawthorne2017OnsetsAF}. This type of ground truth representation contains the pitch activations at different timesteps, which is sufficient for a music transcription model to learn enough to transcribe music with an reasonable transcription accuracy~\cite{Poliner2006ADM, Poliner2007IMPROVINGGF, nam2011classification,trabelsi2017deep, Thickstun2016LearningFO, Thickstun2017InvariancesAD, pedersoli2020improving}. Recently, researchers started introducing sub-tasks such as onset~\cite{Hawthorne2017OnsetsAF} and offset~\cite{kim2019adversarial, kelz2019deep} predictions, and the predictions generated from the sub-tasks are used as input features to another model which produces the final prediction~\cite{Hawthorne2017OnsetsAF, kim2019adversarial}. Choi and Cho~\cite{DBLP:conf/ismir/ChoiC19} have used a reconstruction loss for unsupervised drum transcription, but there is still a lack of evidence on how the reconstruction loss would affect AMT for polyphonic piano music, as spectrograms are seldom utilized to further improve the transcription accuracy. On the contrary, non-neural network AMT models such as graphical models~\cite{berg2014unsupervised}, non-negative matrix factorization~\cite{vincent2009adaptive, smaragdis2003non}, and probabilistic latent component analysis~\cite{benetos2013multiple}, have been successfully trained using solely the reconstruction loss. In view of this, we  explore the possibility of replacing the onset and offset labels with a reconstruction loss because we believe that the piano rolls already contain all necessary information for the model to perform well. We use only the pitch labels in the format of piano rolls and try to reconstruct the spectrogram from the posteriorgram with an L2 reconstruction loss. In this way, we explore how much the model can improve using only piano rolls without handcrafting sub-tasks such as onset and offset predictions. 

\begin{figure*}[t!]
 \centerline{
 \includegraphics[width=\textwidth]{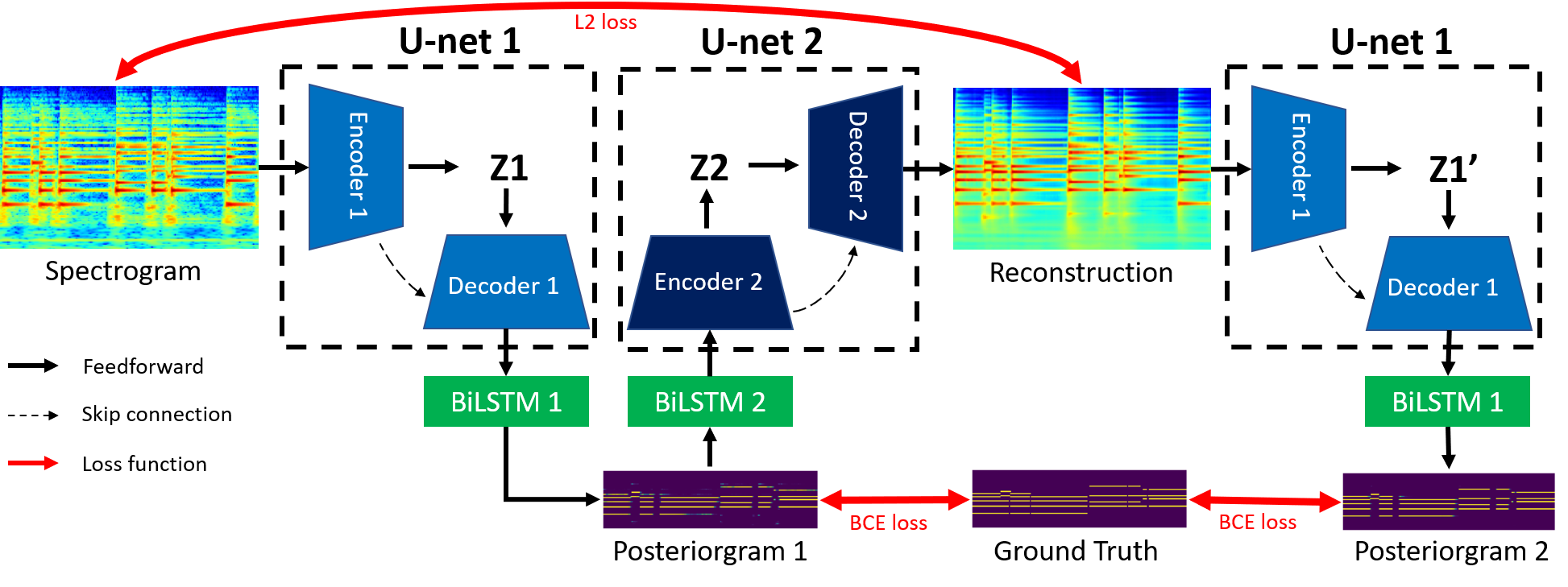}}
 \caption{Proposed Model Architecture. U-net 1 together with BiLSTM 1 forms the transcriber. BiLSTM 2 together with U-net 2 forms the reconstructer. We re-apply the same transcriber to the reconstructed spectrogram to obtain posteriorgram 2, and obtain the final predicted piano roll from this posteriorgram.}
 \label{fig:model}
\end{figure*}

\section{Method}\label{sec:page_size}
\subsection{Formulation}\label{sec: formulation}
We formulate the music transcription problem as a sequence-to-sequence mapping problem. Given an input spectrogram $\textbf{X}_{spec} \in \rm I\!R^{F\times T}$, our model maps it to a posteriorgram $\textbf{Y}_{posterior} \in [0,1]^{P\times T}$, where $F$ is the number of bins on the spectrogram, $P$ is the note index (88 notes, spanning MIDI pitches 21 to 108), and $T$ is the number of time steps. The spectrogram and the posteriorgram share the same number of time steps, which means that a longer input results in a longer output. The piano roll $\textbf{Y}_{roll} \in \{0,1\}^{P\times T}$ can be obtained by applying a hard threshold of 0.5 (same as done by Hawthorne et al.~\cite{Hawthorne2017OnsetsAF}, and Kim and Bello~\cite{kim2019adversarial}) to the posteriorgram, where 0 indicates note off and 1 indicates note on. At the same time, the reconstructor tries to generate the reconstructed spectrogram $\textbf{X}_{recons}\in \rm I\!R^{F\times T}$, using only the posteriorgram.

\subsection{Model Architecture} \label{sec:model}
We adopt the U-netED architecture proposed by Hung et al.~\cite{Hung2019MusicalCS} to be our transcriber (U-net 1 in~\figurename\ref{fig:model}) and our reconstructer (U-net 2), since it has been proven that this model is effective for pitch related tasks. U-net 1 and U-net 2 share the same architecture: four convolution blocks for the encoder part of the U-net, and four convolution blocks for the decoder part of the U-net. Readers can refer to Hung et al.~\cite{Hung2019MusicalCS} for more details on the U-net architecture. U-net 1 and U-net 2 do not share weights. Bidirectional LSTMs (BiLSTMs) are placed after U-net 1 and before U-net 2 to model inter-frame dependencies. The BiLSTM after U-net 1 transforms the outputs of U-net 1 to posteriorgrams. These posteriorgrams are then used to reconstruct the spectrograms via another BiLSTM and U-net 2. There are only skip connections within each U-net, in other words, U-net 2 is forced to reconstruct the spectrograms by using information available from the posteriorgrams. The intuition behind this architecture is that when humans transcribe music, they will typically listen to how their transcription sounds, and compare it to the original audio clip to verify the correctness. The reconstructer (U-net 2) tries to model this procedure. As we will discuss later in Section~\ref{sec:visualize}, the reconstructed spectrogram is a denoised version of the original spectrogram; we then pass the reconstructed spectrogram to U-net 1 again for the final transcription. This is analogous to human transcribers polishing the transcription after having listened to their initial transcription. We obtain our piano roll using this denoised posteriorgram instead of the intermediate posteriorgram, as we show in our experiments that transcriptions based on the second posteriorgram are better than those based on the first one. 
There are three components in the loss to minimize when training our proposed architecture:
\begin{equation}
\begin{split}
    \mathcal{L} = &\text{L2}(\textbf{X}_{spec},\textbf{X}_{recons})+ \sum_{i=1}^{2}\text{BCE}(\textbf{Y}_{post_i},  \textbf{Y}_{roll}) \\
\end{split}
\end{equation}
where $\text{L2}(\cdot, \cdot)$ is the mean square error (L2 loss), $\text{BCE}(\cdot, \cdot)$ is the binary cross-entropy, and $\textbf{Y}_{post_i}$ is the $i$-th posteriorgram for $i \in \{1,2\}$.

One should note that our model formulation is different from Pedersoli et al.~\cite{pedersoli2020improving} despite the fact that they also use U-net as their transcription model. Their model takes as input a spectrogram $\textbf{X}_{spec} \in \rm I\!R^{F\times \tau}$ , where their $\tau$ is much smaller than our timestep size $T$. In addition, their model output covers only one timestep $\textbf{Y}_{posterior} \in [0,1]^{P\times 1}$ as opposed to our model which transcribes the same number of timesteps $T$ as in the spectrogram. Our proposed model is also different from Hawthorne et. al.~\cite{hawthorne2018enabling} despite the fact that both of us are using an encoder-decoder framework. In their work, the main focus is music generation conditioned on piano rolls. They either use the ground truth piano rolls or piano rolls generated from a AMT model~\cite{Hawthorne2017OnsetsAF}. Only when they use the piano roll from the AMT model do they have an encoder part. In addition, the use of reconstruction loss to improve the transcription accuracy is not studied in their work.

\begin{table*}[h]
\centering
 \caption{Results in terms of precision (P), recall (R) and F1-score (F1) for the MAPS dataset. The models reported here use only the pitch labels, and are evaluated using the full recordings in the test set. The transcription accuracy for Sigtia et al.~\cite{Sigtia2015AnEN} and Kelz et al.~\cite{Kelz2016OnTP} are the versions reported in Hawthorne et al.~\cite{Hawthorne2017OnsetsAF}, in which the evaluations were re-run on the full recordings. The 4th row is greyed-out since it uses onset labels. We will not compare out model to it but we put it here as a reference to the readers.}

\small
\resizebox{\textwidth}{!}
{\begin{tabular}{l|ccc|ccc|ccc|}
\cline{2-10}
\multicolumn{1}{c|}{}                             & \multicolumn{3}{c|}{\textbf{Frame}} & \multicolumn{3}{c|}{\textbf{Note}}            & \multicolumn{3}{c|}{\textbf{Note w/ offset}}  \\ \cline{2-10} 
\multicolumn{1}{c|}{}                             & P              & R    & F1          & P             & R             & F1            & P             & R             & F1            \\ \hline
\multicolumn{1}{|l|}{Sigtia et al. 2016~\cite{Sigtia2015AnEN}}          & 72             & \textbf{73.2} & 72.2        & 45            & 49.6          & 46.6          & 17.6          & 19.7          & 18.4          \\ \hline
\multicolumn{1}{|l|}{Kelz et at., 2016~\cite{Kelz2016OnTP}}           & 81.2           & 65.1 & 71.6        & 44.3          & 61.3          & 50.9          & 20.1          & 27.8          & 23.1          \\ \hline
\multicolumn{1}{|l|}{Hawthorne 2018 (Frame only)~\cite{Hawthorne2017OnsetsAF}} & -              & -    & \textbf{76.1}        & -             & -             & 62.7          & -             & -             & 27.9          \\ \hline
\rowcolor{lightgray} 
\multicolumn{1}{|l|}{Hawthorne 2018 (Frame and onset)~\cite{Hawthorne2017OnsetsAF}} & 88.5             & 70.9    & 78.3       & 84.2             & 80.7             & 82.3          & 51.3             & 49.3             & 50.2          \\ \hline

\multicolumn{1}{|l|}{CQT (baseline)} & $79.7 \pm 7.0$ & $67.7 \pm 9.0$ & $72.9 \pm 7.3$ & $57.9 \pm 11.1$ & $57.2 \pm 11.9$ & $57.2 \pm 11.0$ & $34.7 \pm 11.1$ & $34.4 \pm 11.8$ & $34.4 \pm 11.3$  \\ \hline
\multicolumn{1}{|l|}{Mel (baseline)}     & $84.7 \pm 6.0$ & $67.2 \pm 9.7$ & $74.5 \pm 7.2$ & $60.2 \pm 11.3$ & $60.5 \pm 12.2$ & $60.1 \pm 11.2$ & $36.3 \pm 10.7$ & $\textbf{36.7} \pm 11.8$ & $36.3 \pm 11.0$         \\ \hline
\multicolumn{1}{|l|}{CQT (proposed)}     & $86.3 \pm 5.9$ & $61.4 \pm 11.8$ & $71.2 \pm 9.3$ & $67.8 \pm 10.9$ & $57.7 \pm 12.9$ & $61.9 \pm 11.4$ & $39.7 \pm 10.9$ & $34.0 \pm 11.5$ & $36.4 \pm 11.0$           \\ \hline
\multicolumn{1}{|l|}{Mel (proposed)}               & $\textbf{89.3} \pm 5.5$ & $61.9 \pm 11.1$ & $72.5 \pm 8.7$ & $\textbf{71.3} \pm 9.5$ & $\textbf{62.7} \pm 12.5$ & $\textbf{66.3} \pm 10.6$ & $\textbf{41.3} \pm 11.3$ & $36.5 \pm 12.2$ & $\textbf{38.5} \pm 11.6$  \\ \hline
\end{tabular}}
 \label{tab:MAPS}
\end{table*}

\begin{table*}[h]
\centering
\small

 \caption{Results in terms of precision (P), recall (R) and F1-score (F1) for the MAESTRO dataset. Existing work (1st and 2nd rows) on this dataset uses more than just the pitch labels when training their models and hence achieves a much higher accuracy than our model. We will not compare our model to the above mentioned works but we include them here as a reference to the readers.}
\resizebox{\textwidth}{!}
{
\begin{tabular}{l|ccc|ccc|ccc|}
\cline{2-10}
\multicolumn{1}{c|}{}                             & \multicolumn{3}{c|}{\textbf{Frame}} & \multicolumn{3}{c|}{\textbf{Note}}            & \multicolumn{3}{c|}{\textbf{Note w/ offset}}  \\ \cline{2-10} 
\multicolumn{1}{c|}{}                             & P              & R    & F1          & P             & R             & F1            & P             & R             & F1            \\ \hline
\rowcolor{lightgray} 
\multicolumn{1}{|l|}{Hawthorne 2019~\cite{hawthorne2018enabling}} & 92.9                 & 78.5                 & 84.9          & 87.5                 & 85.6                & 86.4            & 66.2                 & 66.8                 &67.4          \\ \hline
\rowcolor{lightgray} 
\multicolumn{1}{|l|}{Kim 2019~\cite{kim2019adversarial}} & 93.1               & {89.8}                 & 91.4          & 98.1                 & 93.2               & 95.6            & 84.1                & 78.1                 & 81.0          \\ \hline
\multicolumn{1}{|l|}{CQT (baseline)} & $91.3 \pm 3.3$ & $65.8 \pm 10.0$ & $76.1 \pm 7.3$ & $68.8 \pm 11.4$ & $65.9 \pm 11.9$ & $67.0 \pm 10.9$ & $36.7 \pm 10.5$ & $35.2 \pm 10.5$ & $35.8 \pm 10.2$          \\ \hline
\multicolumn{1}{|l|}{Mel (baseline)} & $90.2 \pm 3.5$ & $\textbf{71.4} \pm 10.3$ & $\textbf{79.4} \pm 7.1$ & $68.4 \pm 11.6$ & $65.1 \pm 13.2$ & $66.5 \pm 11.9$ & $42.0 \pm 10.7$ & $\textbf{40.1} \pm 11.5$ & $40.9 \pm 10.9$    \\ \hline
\multicolumn{1}{|l|}{CQT (proposed)}            & $89.1 \pm 4.3$ & $67.3 \pm 10.8$ & $76.1 \pm 7.6$ & $72.6 \pm 11.8$ & $63.7 \pm 12.9$ & $67.5 \pm 11.7$ & $44.5 \pm 10.9$ & $39.5 \pm 12.0$ & $41.6 \pm 11.3$      \\ \hline
\multicolumn{1}{|l|}{Mel (proposed)}           & $\textbf{94.0} \pm 2.7$ & $66.1 \pm 12.4$ & $77.0 \pm 9.0$ & $\textbf{78.9} \pm 9.4$ & $\textbf{68.9} \pm 12.6$ & $\textbf{73.3} \pm 10.7$ & $\textbf{44.8} \pm 10.6$ & $39.3 \pm 11.6$ & $\textbf{41.7} \pm 11.0$  \\ \hline
            
\end{tabular}}

 \label{tab:MAESTRO}
\end{table*}

\begin{table*}[h]
\small
\centering
 \caption{Results for the MusicNet dataset in terms of micro average precision ($\mu AP$), accuracy (A), precision (P), recall (R) and F1-score (F1)  Existing models report their results in only frame-level metrics. The reader should note that the frame-level metrics do not directly reflect the note-level metrics in a positive correlation. That is, a high frame-level result  does not guarantee high note-level results, and vice versa.}
 \resizebox{\textwidth}{!}{
\begin{tabular}{l|ccc|ccc|ccc|}
\cline{2-10}
\multicolumn{1}{c|}{}                             & \multicolumn{3}{c|}{\textbf{Frame}} & \multicolumn{3}{c|}{\textbf{Note}}            & \multicolumn{3}{c|}{\textbf{Note w/ offset}}  \\ \cline{2-10} 
\multicolumn{1}{c|}{}                             & $\mu$AP                & A    & F1          & P             & R             & F1            & P             & R             & F1            \\ \hline
\multicolumn{1}{|l|}{Thickstun 2016}          & \textbf{77.3}        & \textbf{55.3}        & -             & -                    & -                    & {-}             & -                    & -                    & {-}       \\ \hline
\multicolumn{1}{|l|}{Pedersoli 2020} & 75.6& -& -&-&-&-&-&-&-\\ \cline{1-10}
\multicolumn{1}{|l|}{CQT (baseline)} & $69.6 \pm 9.4$ & $51.1 \pm 8.3$ & $67.2 \pm 7.6$ & $60.1 \pm 11.7$ & $50.3 \pm 21.4$ & $53.8 \pm 17.4$ & $32.7 \pm 13.4$ & $28.6 \pm 18.1$ & $30.0 \pm 16.2$ \\ \hline
\multicolumn{1}{|l|}{Mel (baseline)} & $71.1 \pm 12.3$ & $53.1 \pm 11.2$ & $\textbf{68.6} \pm 10.1$ & $59.6 \pm 12.3$ & $49.3 \pm 22.3$ & $53.0 \pm 18.4$ & $31.2 \pm 12.1$ & $26.9 \pm 16.6$ & $28.4 \pm 14.8$ \\ \hline
\multicolumn{1}{|l|}{CQT (proposed)}& $71.1 \pm 14.8$ & $42.9 \pm 15.1$ & $58.4 \pm 15.5$ & $\textbf{64.2} \pm \textbf{13.6}$ & $\textbf{51.8} \pm 25.9$ & $55.5 \pm 21.9$ & $\textbf{37.7} \pm \textbf{17.3}$ & $30.4 \pm 21.8$ & $32.5 \pm 20.0$ \\ \hline
\multicolumn{1}{|l|}{Mel (proposed)}& $71.1 \pm 14.6$ & $50.7 \pm 12.6$ & $66.3 \pm 11.9$ & $63.5 \pm 14.0$ & $51.4 \pm 23.9$ & $\textbf{55.7} \pm 20.2$ & $37.7 \pm 18.8$ & $\textbf{32.2} \pm 23.2$ & $\textbf{34.1} \pm 21.6$ \\ \hline
 
\end{tabular}}
 \label{tab:MusicNet}
\end{table*}

\section{Experimental Setup}\label{sec:experiment}
\subsection{Dataset}
We evaluate our proposed model using three different datasets, MAPS~\cite{emiya2010maps}, MAESTRO~\cite{hawthorne2018enabling}, and MusicNet~\cite{Thickstun2016LearningFO} to verify the effectiveness of our approach. For the MAPS dataset, we follow the same train and test splits as in the existing literature~\cite{Sigtia2015AnEN,Hawthorne2017OnsetsAF} by removing overlapping songs in the train set that are also present in the test set. For the MAESTRO dataset, we use the train split provided to train our model, and we use the test split to evaluate our performance. For the MusicNet dataset, the recordings with IDs: 2303, 2382, 1819 are reserved for testing, and the rest are used for training, following the same setup as in the literature~\cite{Thickstun2016LearningFO, Thickstun2017InvariancesAD, pedersoli2020improving}. All audio recordings from the datasets are downsampled to $16$~kHz.

\subsection{Training Procedure and Hyperparameters}
During training, a segment of $327,680$ samples (roughly 20 seconds) is extracted from each song with a random starting point in each epoch to train the model. We convert the audio segments to spectrograms on-the-fly using nnAudio~\cite{cheuk2019nnaudio}, a tool for GPU-based spectrogram extraction in PyTorch. We experiment with both Constant-Q transform (CQT) and Mel spectrograms as the input representations. For CQT, the starting bin is at $27.5$~Hz which corresponds to MIDI pitch 21. There are 176 bins in the CQT, with 24 bins per octave, covering exactly the range of a piano. For the Mel spectrogram, we keep the parameters the same as in Hawthorne et al.~\cite{Hawthorne2017OnsetsAF} so we can make a fair comparison later. The window length is 2,048; 229 Mel filter banks are used; and the starting frequency is $30$~Hz. For both spectrograms, the hop length is set to 512. The magnitude of the spectrograms is compressed with a $\log$ function and then normalized using min-max normalization.
We train out model for 2,000 epochs as the model converges at this point for all datasets. For each epoch, a different segment with a random starting position is extracted from each recording to prevent over-fitting. This method has proven to be effective in existing studies~\cite{Hawthorne2017OnsetsAF, kim2019adversarial}. The batch size is 32 for all our experiments.
Adam~\cite{kingma2014adam} with decaying learning rate is used as the optimizer. The initial learning rate is $6\times 10^{-4}$, and the learning rate decays with the factor of $0.98$ every $10,000$ steps.

All components of the neural network (Both U-nets and the BiLSTMs) are trained simultaneously, i.e. the three losses mentioned in Section~\ref{sec:model} are minimized at the same time. The same U-net 1 is used to transcribe both the original and reconstructed spectrogram. In order to evaluate the effectiveness of the reconstruction loss, we remove the reconstruction part of the network, leaving behind only U-net 1 and BiLSTM1, and use this model as the \textbf{baseline model}. The results of our experiments are presented in Section~\ref{sec:Results}.

\subsection{Evaluation Method}\label{subsec:eval}
We evaluate our model's performance using the metrics used in the MIREX multi-F0 and note tracking evaluations~\cite{bay2009evaluation}, which are implemented in \textbf{mir\_eval}~\cite{Raffel2014MIR_EVALAT}. In order to make this study comprehensive, we report not only frame-level metrics, but also onset-only note-level metrics, and onset-offset note-level metrics. The onset tolerance is 50msec and the offset tolerance is 20\% of the duration or 50msec, whichever is larger. Each frame in our piano roll is equivalent to $3.125$~ms of the audio recording. Moreover, as pointed out by Hawthorne et al.~\cite{Hawthorne2017OnsetsAF}, evaluating on only 30 seconds of each audio file in the test set is not representative. Therefore, we evaluate our model using the entire audio recordings in the test set and obtain the precision, recall and F1-score, and report the average of these scores.

For MusicNet, the existing literature reports only the micro average-precision ($\mu AP$) and frame accuracy~\cite{Thickstun2016LearningFO, Thickstun2017InvariancesAD, pedersoli2020improving}. When reporting results using MusicNet, we have modified our frame-level evaluation to match with the metrics reported in the literature, however we still keep reporting the same note-level metrics as in the MAPS and MAESTRO evaluations.


\section{Results}\label{sec:Results}

\begin{figure*}[!htbp]
 \centerline{
 \includegraphics[width=\textwidth]{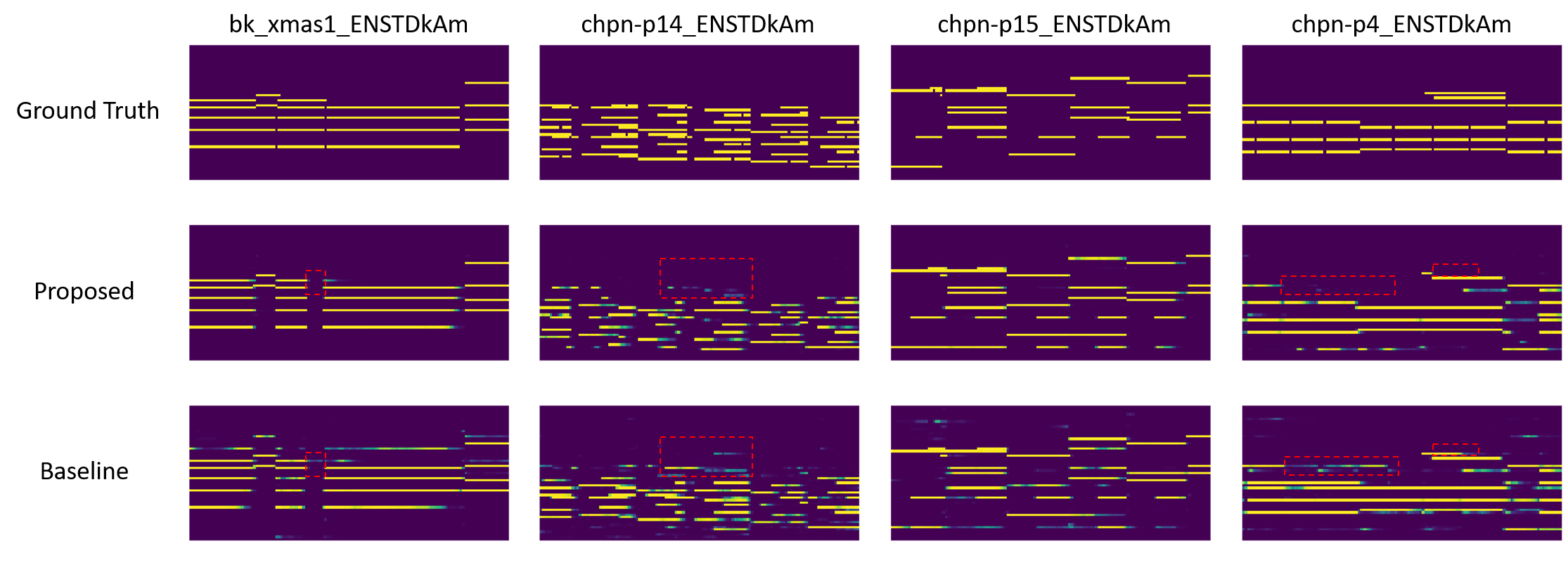}}
 \caption{Four examples from the test set of the MAPS dataset are used to showcase the behaviour of our proposed model. Our model does not make a prediction when there is too much uncertainty (as indicated by the red dashed squares), which results in a higher precision and lower recall. The piano rolls shown here are obtained with the model using Mel spectrograms, the CQT based models have the same behaviour.}
 \label{fig:pianoroll}
\end{figure*}

We evaluate our models using three different datasets. Our results indicate that, in general,  when using the Mel spectrogram as the input representation to the model, it performs better than using the CQT spectrogram as the input. A relatively inferior model performance using CQT has also been reported in existing literature~\cite{cheuk2020impact, Kelz2016OnTP, wang2018singing, Hawthorne2017OnsetsAF}, which is confirmed once again in our study. In the following subsections, we will discuss the results for each dataset in detail. Readers can refer to Section~\ref{subsec:eval} for details on the evaluation metrics.

\subsection{MAPS dataset}\label{subsec: MAPS result}
Our results on the MAPS dataset are presented in Table~\ref{tab:MAPS}. It can be seen from the table that our proposed model with Mel spectrograms outperforms the existing models which use only the piano roll as the ground truth (row 1 to 3) in terms of note-level metrics. The F1-score for the note metrics with onsets has been improved by $19.7$ and $15.4$ percentage points relative to Sigtia's~\cite{Sigtia2015AnEN} and Kelz's~\cite{Kelz2016OnTP} result respectively. For note metrics with offsets, our model outperforms Sigtia and Kelz's model by $20.1$ and $15.4$ percentage points respectively. When we use only the frame module from Hawthorne et al.~\cite{Hawthorne2017OnsetsAF}, their model performance decays a lot, and becomes worse than our proposed model. We outperform their model in the `note with offset' metrics by a large margin when the model has no explicit access to the onset labels. However, our model is still unable to outperform the state-of-the-art model (row 4), in which they introduced a supervised onset prediction sub-task, while our model uses an unsupervised spectrogram reconstruction loss. Still, this reconstruction loss is able to boost the frame-level precision to a state-of-the-art performance.

To study the impact of the reconstruction loss, we use our proposed model without the reconstruction module as our baseline model. The reconstruction loss successfully improves the onset-only note-level metrics and the `note with offset' metrics. The Wilcoxon signed-rank test on the recording-level F1-scores shows that the improvements due to our proposed model are statistically significant when compared to the baseline model in these two metrics (p-value of 0). The F1-score for the frame metrics of our proposed model is worse than the baseline model (i.e. without the reconstruction module). This is, however, due to a huge decay in recall, the precision actually improves when using the reconstruction module. A similar pattern can be seen for the `note with offset' metrics too. It appears that adding the reconstruction module enables our model to discover the onset features from the spectrogram and hence improving the onset-only note-level metrics. In addition, it also makes the model more conservative judging from the decay in the recall under the frame and `note with offset' metrics. A similar pattern can be observed when CQT is used as the input representation instead of Mel spectrograms. The piano rolls predicted by our proposed and baseline models are shown in \figurename~\ref{fig:pianoroll}. It can be seen from the figure that the proposed model produces less fragmented note predictions when compared to the baseline model without the reconstruction component. When our proposed model is too uncertain about the note prediction, it omits the predictions as indicated by the red dashed squares, which leads to a decrease of the recall metric when compared to the baseline model.

\subsection{MAESTRO}
The total duration and total number of notes in the MAESTRO dataset is 10 times more than in the MAPS dataset. Heavily supervised models which use pitch, onset and even offset labels~\cite{hawthorne2018enabling, kim2019adversarial} benefit a lot from this huge dataset. Our model still benefits from using this dataset, but not in the same degree as the heavily supervised models. The improvement can mostly be seen in the note-level metrics, for example our proposed model has a boost of $7.0$ percentage points in F1-score for the note-level metrics from 66.3\% (trained with the MAPS dataset) to 73.3\% (trained with the MAESTRO dataset) when using Mel spectrogram as the input representation. There is also a slightly higher F1-score for the frame-level metrics, from $72.5\%$ to $77\%$, finally the `note with offset' metrics also increase from 38.5\% to 41.7\%. Again, the Wilcoxon signed-rank test shows that the improvements are statistically significant for note-level and `note with offset' metrics (p-value of 0). Similar to what we have observed when training on the MAPS dataset, our model's predictions become more conservative when we include the reconstruction module, resulting in a decay in recall for the frame-level but a boost in frame-level precision.
To the authors' knowledge, there are no reports of training with the MAESTRO dataset using only the piano roll ground truth, at the time of writing this paper. Many models include onset labels~\cite{hawthorne2018enabling} or even offset labels~\cite{kim2019adversarial} when training on this dataset, therefore the readers should note that the comparison made in Table~\ref{tab:MAESTRO} is not entirely fair. The models using onset and offset labels are of course much better than our model. Nonetheless, our proposed model is still able to outperform them with the frame-level precision. When comparing our proposed model with the baseline model, a clear improvement on the note-level metrics can be seen. Again, the model with Mel Spectrogram as input performs the best. 
\vspace{-1mm}
\subsection{MusicNet dataset}

MusicNet should be the most difficult dataset among the three because of its wide range of musical instruments. Existing literature using this dataset reports only frame-level metrics, but we will still compare our model with them.

When it comes to the average Precision ($\mu AP$), the models of Thickstun et al.~\cite{Thickstun2017InvariancesAD} and Pedersoli et al.~\cite{pedersoli2020improving} outperform our model. When it comes to frame accuracy, our proposed model performs slightly less good than Thickstun's model. However, as pointed out by Hawthorne et al.~\cite{Hawthorne2017OnsetsAF}, the frame-level metrics alone might not provide the full picture of the model performance, it is therefore necessary to include note-level metrics or even `note with offset' metrics in order to have a more comprehensive evaluation. Unfortunately, these metrics together with the frame-level precision are not reported in existing publications, and we are unable to verify if our model also outperform others in this case.

As we can see, both of our proposed models (using CQT and Mel spectrograms as the input representation) and one baseline model (Mel) reach the same muAP of $71.1\%$, but the note-level metrics provide a different picture. Therefore reporting only the frame-level metrics, or even one sub metric under the frame-level metrics could be misleading. Similar to MAPS and MAESTRO datasets, our proposed model improves on both  note-level metrics. Overall, our proposed model behaves similarly regardless of which dataset is being used: 1) Improvement in the note-level metrics; 2) Improvement in the frame-level precision metrics; 3) Small decay over the frame-level recall metrics. Since there are only three test samples, we are unable to conduct a statistical
significance analysis for this dataset.

\subsection{Features Learned by the U-net}\label{sec:visualize}

\begin{figure*}[!htbp]
 \centerline{
 \includegraphics[width=\textwidth]{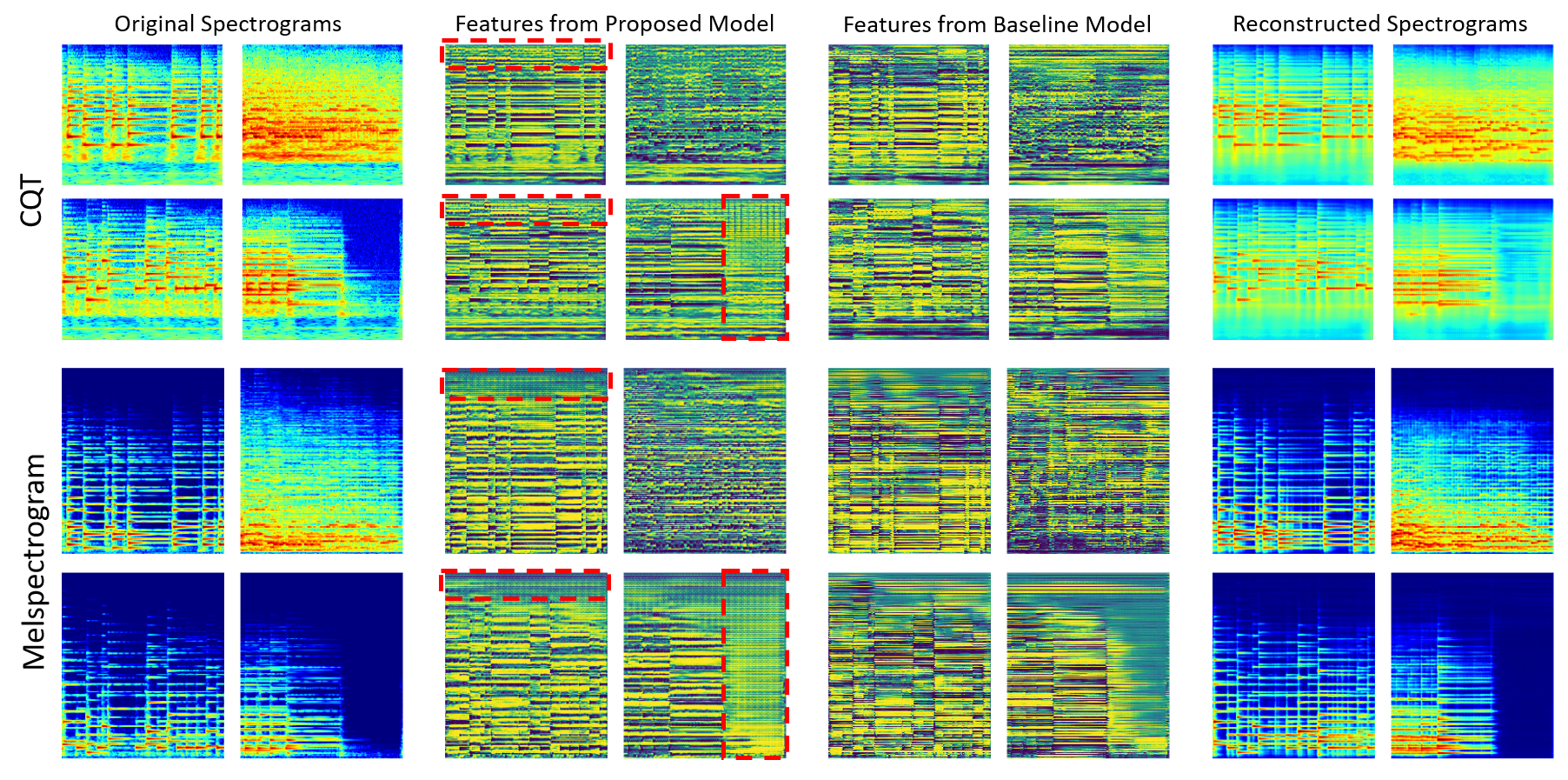}}
 \caption{First column: the original spectrograms that are fed to our models. Second and third columns: the output from U-net 1 with (proposed model) and without the reconstruction loss (baseline model). Fourth column: the reconstructed spectrograms generated by U-net 2 in our proposed model. The dashed red rectangle indicates the grid structures learned when using the reconstruction loss.}
 \label{fig:visualize}
\end{figure*}

The evaluation results show that our proposed model is able to improve in terms of note-level metrics; now we would like to discuss the reason behind it. The output of U-net 1 (before the first BiLSTM) is shown in \figurename~\ref{fig:visualize}. For both the CQT and Mel spectrograms, the output of U-net 1 (before the BiLSTM) clearly contains the pitch information of the original spectrogram, this can be seen as the horizontal lines in the figure. In the case of CQT, the pitch information is nearly the same as in the original spectrogram.  In this new feature, when the value is 0 (dark color), it means the note is activated, when the value is close to 1 (yellow color), it means there is no event happening. For our proposed model, however, when the recording ends and no more notes appear, the output of U-net 1 is not plain yellow. Instead, a grid-like structure is found in these locations. We suspect that our U-net 1 is trying to learn how to count the beats similar to what Ycart and Benetos have reported~\cite{ycart2017study}. Using this beat counting feature as an input to BiLSTM 1 could be beneficial to onset detection and hence explain the improved note-level metrics for transcription. For our baseline model, where there is no reconstruction module, the grid-like structure is not as strong as the model with reconstruction loss (although one can still find some traces of this grid-like structure when they magnify the feature map and examine the pixels carefully). We believe that when we let the model reconstruct the spectrogram from the posteriorgram, the onset detection or even offset detection are learned in an unsupervised manner.

This observation raises the question if music transcription can be learned via only training the model with spectrogram reconstruction. We examined this idea by removing the binary cross-entropy loss between the posteriorgrams and the ground truth piano roll, and train the model with only the L2 reconstruction loss. Unfortunately, the model failed to learn anything meaningful in the posteriorgram, despite the fact that the reconstructed spectrograms become nearly a perfect reconstruction of the original spectrogram. This is probably due to the fact that when the constraints on the posteriorgram are removed, the model deviates from the transcription task and tries to carry out a reconstruction as good as it can. In order to train the model in a completely unsupervised manner, we need some constraints for the posteriorgram when we stop providing the ground truth piano roll to the model. 

\section{Conclusion}
In this paper, we deviate from the current trend in automatic music transcription research by training our model with only the piano roll, without providing explicit onset and offset labels to our proposed model using U-nets and BiLSTMs. Instead, we add a reconstruction module in our model so that it reconstructs the (noise reduced) spectrograms from the posteriorgrams using the L2 loss. Applying the transcriber on the reconstructed spectrograms yields a higher frame-level precision and note-level transcription accuracy. We also discovered that the model is able to learn a beat counting-like feature, which  helps to improve the transcription at the note-level.
We evaluated our models on a music transcription tasks using three major datasets, MAPS, MAESTRO, and MusicNet. Our proposed model improves the note-level metrics for all three of these datasets, when compared to our baseline model without the reconstruction loss, and outperforms existing models which use only the ground truth piano roll. The reconstruction loss also boosts the frame-level precision to the state-of-the-art, proving that spectrogram reconstruction is a useful task for music transcription.

To extend this work, one could study the effect of reconstruction loss on a model which has been trained with onset and offset labels. Would the reconstruction loss further improve the transcription accuracy? Or would the model just ignore the reconstruction due to the fact that onset and offset labels are much stronger constrains? In future research, it is also worth to look for new loss functions or constrains that can improve the recall together with the precision without sacrificing either of them. This paper serves as a starting point for researchers who want to further explore semi-supervised music transcription models or even fully unsupervised music transcription models. Possible research questions leading to a fully unsupervised model are (1) Could we find a custom unsupervised loss that enables the model to improve music transcription accuracy without providing piano rolls as the labels for training? (2) What constrains can we apply to the posteriorgram (e.g. sparsity loss) that could prevent the model from straying away from the transcription task when the ground truth labels are absent.

\section{Acknowledgements}

This work is supported by Singapore International Graduate Award (SINGA) provided by the Agency for Science, Technology and Research (A*STAR) under grant no. SING-2018-02-0204, MOE Tier 2 grant no. MOE2018-T2-2-161, and SRG ISTD 2017 129. 


%
\IEEEpeerreviewmaketitle

\bibliographystyle{IEEEtran}
\bibliography{IEEEfull}

\begin{thebibliography}{10}
\providecommand{\url}[1]{#1}
\csname url@samestyle\endcsname
\providecommand{\newblock}{\relax}
\providecommand{\bibinfo}[2]{#2}
\providecommand{\BIBentrySTDinterwordspacing}{\spaceskip=0pt\relax}
\providecommand{\BIBentryALTinterwordstretchfactor}{4}
\providecommand{\BIBentryALTinterwordspacing}{\spaceskip=\fontdimen2\font plus
\BIBentryALTinterwordstretchfactor\fontdimen3\font minus
  \fontdimen4\font\relax}
\providecommand{\BIBforeignlanguage}[2]{{%
\expandafter\ifx\csname l@#1\endcsname\relax
\typeout{** WARNING: IEEEtran.bst: No hyphenation pattern has been}%
\typeout{** loaded for the language `#1'. Using the pattern for}%
\typeout{** the default language instead.}%
\else
\language=\csname l@#1\endcsname
\fi
#2}}
\providecommand{\BIBdecl}{\relax}
\BIBdecl

\bibitem{cemgil2004bayesian}
A.~T. Cemgil, \emph{Bayesian music transcription}.\hskip 1em plus 0.5em minus
  0.4em\relax Ph.D. thesis, Radbound University Nijmegen Netherlands, 2004.

\bibitem{carvalho2017towards}
R.~G.~C. Carvalho and P.~Smaragdis, ``Towards end-to-end polyphonic music
  transcription: Transforming music audio directly to a score,'' in \emph{2017
  IEEE Workshop on Applications of Signal Processing to Audio and Acoustics
  (WASPAA)}.\hskip 1em plus 0.5em minus 0.4em\relax IEEE, 2017, pp. 151--155.

\bibitem{roman2018end}
M.~A. Rom{\'{a}}n, A.~Pertusa, and J.~Calvo{-}Zaragoza, ``An end-to-end
  framework for audio-to-score music transcription on monophonic excerpts.'' in
  \emph{ISMIR}, 2018, pp. 34--41.

\bibitem{roman2019holistic}
------, ``A holistic approach to polyphonic music transcription with neural
  networks,'' in \emph{Proceedings of the 20th International Society for Music
  Information Retrieval Conference, {ISMIR} 2019, Delft, The Netherlands,
  November 4-8, 2019}, 2019, pp. 731--737.

\bibitem{cuthbert2010music21}
M.~S. Cuthbert and C.~Ariza, ``music21: A toolkit for computer-aided musicology
  and symbolic music data,'' 2010.

\bibitem{huang2020pop}
Y.-S. Huang and Y.-H. Yang, ``Pop music transformer: Generating music with
  rhythm and harmony,'' \emph{arXiv preprint arXiv:2002.00212}, 2020.

\bibitem{magalhaeschordify}
J.~P. Magalhaes, ``Chordify: Three years after the launch,'' 2015.

\bibitem{Poliner2006ADM}
G.~E. Poliner and D.~P.~W. Ellis, ``A discriminative model for polyphonic piano
  transcription,'' \emph{EURASIP Journal on Advances in Signal Processing},
  vol. 2007, pp. 1--9, 2006.

\bibitem{Poliner2007IMPROVINGGF}
------, ``Improving generalization for classification-based polyphonic piano
  transcription,'' \emph{2007 IEEE Workshop on Applications of Signal
  Processing to Audio and Acoustics}, pp. 86--89, 2007.

\bibitem{BoulangerLewandowski2012ModelingTD}
N.~Boulanger-Lewandowski, Y.~Bengio, and P.~Vincent, ``Modeling temporal
  dependencies in high-dimensional sequences: Application to polyphonic music
  generation and transcription,'' in \emph{ICML}, 2012.

\bibitem{springenberg2014striving}
J.~T. Springenberg, A.~Dosovitskiy, T.~Brox, and M.~Riedmiller, ``Striving for
  simplicity: The all convolutional net,'' \emph{arXiv preprint
  arXiv:1412.6806}, 2014.

\bibitem{Sigtia2015AnEN}
S.~Sigtia, E.~Benetos, and S.~Dixon, ``An end-to-end neural network for
  polyphonic piano music transcription,'' \emph{IEEE/ACM Transactions on Audio,
  Speech, and Language Processing}, vol.~24, pp. 927--939, 2015.

\bibitem{ullrich2018music}
K.~Ullrich and E.~van~der Wel, ``Music transcription with convolutional
  sequence-to-sequence models,'' 2018.

\bibitem{Hawthorne2017OnsetsAF}
C.~Hawthorne, E.~Elsen, J.~Song, A.~Roberts, I.~Simon, C.~Raffel, J.~Engel,
  S.~Oore, and D.~Eck, ``Onsets and frames: Dual-objective piano
  transcription,'' in \emph{ISMIR}, 2017.

\bibitem{kim2019adversarial}
J.~W. Kim and J.~P. Bello, ``Adversarial learning for improved onsets and
  frames music transcription,'' \emph{International Society forMusic
  Information Retrieval Conference}, pp. 670--677, 2019.

\bibitem{kelz2019deep}
R.~Kelz, S.~B{\"o}ck, and G.~Widmer, ``Deep polyphonic adsr piano note
  transcription,'' in \emph{ICASSP 2019-2019 IEEE International Conference on
  Acoustics, Speech and Signal Processing (ICASSP)}.\hskip 1em plus 0.5em minus
  0.4em\relax IEEE, 2019, pp. 246--250.

\bibitem{kameoka2007multipitch}
H.~Kameoka, T.~Nishimoto, and S.~Sagayama, ``A multipitch analyzer based on
  harmonic temporal structured clustering,'' \emph{IEEE Transactions on Audio,
  Speech, and Language Processing}, vol.~15, no.~3, pp. 982--994, 2007.

\bibitem{berg2014unsupervised}
T.~Berg-Kirkpatrick, J.~Andreas, and D.~Klein, ``Unsupervised transcription of
  piano music,'' in \emph{Advances in neural information processing systems},
  2014, pp. 1538--1546.

\bibitem{DBLP:conf/ismir/ChoiC19}
K.~Choi and K.~Cho, ``Deep unsupervised drum transcription,'' in \emph{ISMIR
  4-8, 2019}, A.~Flexer, G.~Peeters, J.~Urbano, and A.~Volk, Eds., 2019, pp.
  183--191.

\bibitem{ycart2019blending}
A.~Ycart, A.~McLeod, E.~Benetos, K.~Yoshii \emph{et~al.}, ``Blending acoustic
  and language model predictions for automatic music transcription.''\hskip 1em
  plus 0.5em minus 0.4em\relax ISMIR, 2019.

\bibitem{ronneberger2015u}
O.~Ronneberger, P.~Fischer, and T.~Brox, ``U-net: Convolutional networks for
  biomedical image segmentation,'' in \emph{International Conference on Medical
  image computing and computer-assisted intervention}.\hskip 1em plus 0.5em
  minus 0.4em\relax Springer, 2015, pp. 234--241.

\bibitem{emiya2010maps}
V.~Emiya, N.~Bertin, B.~David, and R.~Badeau, ``Maps-a piano database for
  multipitch estimation and automatic transcription of music,'' 2010.

\bibitem{hawthorne2018enabling}
C.~Hawthorne, A.~Stasyuk, A.~Roberts, I.~Simon, C.-Z.~A. Huang, S.~Dieleman,
  E.~Elsen, J.~Engel, and D.~Eck, ``Enabling factorized piano music modeling
  and generation with the {MAESTRO} dataset,'' in \emph{International
  Conference on Learning Representations}, 2019.

\bibitem{Thickstun2017InvariancesAD}
J.~Thickstun, Z.~Harchaoui, D.~P. Foster, and S.~M. Kakade, ``Invariances and
  data augmentation for supervised music transcription,'' \emph{2018 IEEE
  International Conference on Acoustics, Speech and Signal Processing
  (ICASSP)}, pp. 2241--2245, 2017.

\bibitem{nam2011classification}
J.~Nam, J.~Ngiam, H.~Lee, M.~Slaney \emph{et~al.}, ``A classification-based
  polyphonic piano transcription approach using learned feature
  representations.'' in \emph{ISMIR}, 2011, pp. 175--180.

\bibitem{trabelsi2017deep}
C.~Trabelsi, O.~Bilaniuk, Y.~Zhang, D.~Serdyuk, S.~Subramanian, J.~F. Santos,
  S.~Mehri, N.~Rostamzadeh, Y.~Bengio, and C.~J. Pal, ``Deep complex
  networks,'' \emph{arXiv preprint arXiv:1705.09792}, 2017.

\bibitem{Thickstun2016LearningFO}
J.~Thickstun, Z.~Harchaoui, and S.~M. Kakade, ``Learning features of music from
  scratch,'' \emph{ArXiv}, vol. abs/1611.09827, 2016.

\bibitem{pedersoli2020improving}
F.~Pedersoli, G.~Tzanetakis, and K.~M. Yi, ``Improving music transcription by
  pre-stacking a u-net,'' in \emph{ICASSP 2020-2020 IEEE International
  Conference on Acoustics, Speech and Signal Processing (ICASSP)}.\hskip 1em
  plus 0.5em minus 0.4em\relax IEEE, 2020, pp. 506--510.

\bibitem{bock2012polyphonic}
S.~B{\"o}ck and M.~Schedl, ``Polyphonic piano note transcription with recurrent
  neural networks,'' in \emph{2012 IEEE international conference on acoustics,
  speech and signal processing (ICASSP)}.\hskip 1em plus 0.5em minus
  0.4em\relax IEEE, 2012, pp. 121--124.

\bibitem{vincent2009adaptive}
E.~Vincent, N.~Bertin, and R.~Badeau, ``Adaptive harmonic spectral
  decomposition for multiple pitch estimation,'' \emph{IEEE Transactions on
  Audio, Speech, and Language Processing}, vol.~18, no.~3, pp. 528--537, 2009.

\bibitem{smaragdis2003non}
P.~Smaragdis and J.~C. Brown, ``Non-negative matrix factorization for
  polyphonic music transcription,'' in \emph{2003 IEEE Workshop on Applications
  of Signal Processing to Audio and Acoustics (IEEE Cat. No. 03TH8684)}.\hskip
  1em plus 0.5em minus 0.4em\relax IEEE, 2003, pp. 177--180.

\bibitem{benetos2013multiple}
E.~Benetos and S.~Dixon, ``Multiple-instrument polyphonic music transcription
  using a temporally constrained shift-invariant model,'' \emph{The Journal of
  the Acoustical Society of America}, vol. 133, no.~3, pp. 1727--1741, 2013.

\bibitem{Hung2019MusicalCS}
Y.-N. Hung, I.~P. Chiang, Y.~Chen, and Y.-H. Yang, ``Musical composition style
  transfer via disentangled timbre representations,'' in \emph{IJCAI}, 2019,
  pp. 4697--4703.

\bibitem{Kelz2016OnTP}
R.~Kelz, M.~Dorfer, F.~Korzeniowski, S.~B{\"o}ck, A.~Arzt, and G.~Widmer, ``On
  the potential of simple framewise approaches to piano transcription,'' in
  \emph{ISMIR}, 2016.

\bibitem{cheuk2019nnaudio}
K.~W. Cheuk, H.~Anderson, K.~Agres, and D.~Herremans, ``nnaudio: An on-the-fly
  gpu audio to spectrogram conversion toolbox using 1d convolution neural
  networks,'' \emph{arXiv preprint arXiv:1912.12055}, 2019.

\bibitem{kingma2014adam}
D.~P. Kingma and J.~Ba, ``Adam: A method for stochastic optimization,''
  \emph{arXiv preprint arXiv:1412.6980}, 2014.

\bibitem{bay2009evaluation}
M.~Bay, A.~F. Ehmann, and J.~S. Downie, ``Evaluation of multiple-f0 estimation
  and tracking systems.'' in \emph{ISMIR}, 2009, pp. 315--320.

\bibitem{Raffel2014MIR_EVALAT}
C.~Raffel, B.~McFee, E.~J. Humphrey, J.~Salamon, O.~Nieto, D.~Liang, and
  D.~P.~W. Ellis, ``Mir\_eval: A transparent implementation of common mir
  metrics,'' in \emph{ISMIR}, 2014.

\bibitem{cheuk2020impact}
K.~W. Cheuk, K.~Agres, and D.~Herremans, ``The impact of audio input
  representations on neural network based music transcription,''
  \emph{International Joint Conference on Neural Networks}, 2020.

\bibitem{wang2018singing}
C.-i. Wang and G.~Tzanetakis, ``Singing style investigation by residual siamese
  convolutional neural networks,'' in \emph{2018 IEEE International Conference
  on Acoustics, Speech and Signal Processing (ICASSP)}.\hskip 1em plus 0.5em
  minus 0.4em\relax IEEE, 2018, pp. 116--120.

\bibitem{ycart2017study}
A.~Ycart, E.~Benetos \emph{et~al.}, ``A study on lstm networks for polyphonic
  music sequence modelling,'' in \emph{ISMIR}, 2017.

\end{thebibliography}
%



\end{document}